\documentclass[10pt,preprint]{aastex}

\usepackage{graphicx}
\usepackage{tabularx}
\usepackage{subfigure}
\usepackage{color}
\usepackage{float}
\usepackage{natbib}
\usepackage{mathbbol}
\usepackage{amssymb}
\usepackage{amsmath}
\newcommand{\be}{\begin{equation}}
\newcommand{\ee}{\end{equation}}

\newcommand{\ba}{\begin{eqnarray}}
\newcommand{\ea}{\end{eqnarray}}

\renewcommand{\v}{{\bf v}}

\newcommand\eg{\textit{e.g.,\ }}

\newcommand{\Bf}{{magnetic field}}

\newcommand{\Ef}{{electric  field}}

\newcommand{\ms}{magnetosphere}

\newcommand{\Fermi}{{\it Fermi}}

\begin{document}

\title{High energy tails of pulsar gamma-ray emission}

\author{
G.~T.~Richards\altaffilmark{1},
M.~Lyutikov\altaffilmark{2},
}
\altaffiltext{1}{School of Physics and Center for Relativistic Astrophysics, Georgia Institute of Technology, 837 
State Street NW, Atlanta, GA 30332-0430}
\altaffiltext{2}{Department of Physics and Astronomy, Purdue University, West Lafayette, IN 47907, USA  and Department of Physics and McGill Space Institute, McGill University, 3600 University Street, Montreal, Quebec H3A 2T8, Canada}

\begin{abstract}
We perform spectral analyses of four bright $\gamma$-ray pulsars: PSR J0007+7303, Vela,
Geminga and  PSR J2021+3651 concentrating on the high-energy tails, defined  as emission
above 10\,GeV. The two competing models of pulsar $\gamma$-ray emission predict
qualitatively different spectra well above the break energy: curvature emission predicts
an exponential cut-off in the spectra, while the inverse-Compton scattering
mechanism favors a power-law. We perform fits to the phase-averaged spectral energy
distributions for each of the four pulsars. We find that in all cases both models fit the data equally well---the present data set
 does not allow any firm claim to be made about the shape of the
spectra above 10\,GeV.
 In no case is the power-law fit or exponential cut-off fit
significantly preferred over the other. 
 The Crab pulsar remains the only known pulsar in
which the power-law fit is clearly preferred over the exponential cut-off.
% is this still true after the recent Vela developments -- HESS?
% According to Nepomuk -- yes, the HESS results don't allow any more spectral info than
% the Fermi-LAT stuff...
\end{abstract}


\section{Introduction}

%\section{\bf PULSAR VERY HIGH ENERGY $\gamma$-RAY EMISSION}
 
 At present more than 200
pulsars\footnote{\color{blue}\detokenize{
https://confluence.slac.stanford.edu/display/GLAMCOG/Public+List+of+LAT-Detected+Gamma-Ray+Pulsars}} are
known to emit $\gamma$-rays in the high-energy
(HE; E $\gtrsim$ 100\,MeV) band, and this number is
steadily growing~\citep{2013ApJS..208...17A}.    %It has been realized that  a
%considerable fraction of young pulsars emit  $\gamma$-rays.
%not sure of point mentioning young pulsars here -gtr 
Understanding the emission mechanism of $\gamma$-ray pulsars and its relationship to radio
pulsar emission is one of the most pressing questions in high-energy astrophysics.
 
Since the launch of the \Fermi\ satellite, the $\gamma$-ray data unambiguously point to an
outer-magnetospheric origin of the high-energy emission.  The observed spectra lack the
super-exponential cut-off expected in ``polar cap'' models~\citep{2008Sci...322.1221A}, and there are far
more radio-quiet $\gamma$-ray pulsars than are predicted in the polar cap scenario. There
are a few independent {\it geometrical} models of pulsar high-energy 
emission that  assume enhanced particle acceleration and radiation along the last open
field
lines of a magnetic dipole~\citep{1995ApJ...438..314R, 2010ApJ...715.1282B}.
Though {\it geometrical} models of the $\gamma$-ray emission from pulsars have been quite
successful, the underlying physics of particle acceleration and emission remains an open
question.  In modeling  the electrodynamics of the pulsar gap, it is
typically assumed that the primary emission mechanism  is curvature
radiation~\citep[\eg][]{1996ApJ...470..469R,Baring99}. 
% need to fill missing citation here.

Our current theoretical models of pulsar radiation are not yet able
 to produce solid  predictions.  Indeed, there are several unanswered questions: 
the global kinetic structure of the magnetosphere, the location of the gaps, 
the current closure, and the location of pair production \citep{2014ApJ...795L..22C,2015MNRAS.448..606C,2015MNRAS.449.2759B}.  Also, the process
is non-local---photons emitted at one place can produce an avalanche on different field
lines, so the gap may  not self-close even though it produces a cascade.  
  
The VERITAS Collaboration  detected  pulsed emission from the Crab pulsar at energies above
100\,GeV~\citep{2011Sci...334...69V}. This  discovery was confirmed by MAGIC,
who have recently claimed the detection of the pulsed emission up to
1.5\,TeV~\citep{2016A&A...585A.133A}. Most importantly, the
photon spectrum of the pulsed emission between 100\,MeV and 400\,GeV is best described by
a broken power law and is statistically  preferred over a power law with an
exponential cut-off. 
 
 Conventionally,   $\gamma$-ray emission from pulsars is attributed to incoherent
curvature radiation. 
The detection of the Crab pulsar by VERITAS~\citep{2011Sci...334...69V}  implies the
importance of inverse-Compton (IC) scattering for the production of $\gamma$-ray photons.
This follows from a very general argument for an upper limit %, independent of the
%particular details of
%the acceleration mechanism,  
on the possible energies  of curvature
photons~\citep[following a similar approach applicable to synchrotron
emission;][]{1996ApJ...457..253D,2010MNRAS.405.1809L}. 
Assuming  that the  radius of curvature $R_c $ of the  \Bf\ lines  is a fraction  $\xi$ of the light cylinder radius $R_c = \xi R_{LC}$ and 
balancing radiative losses with acceleration by the \Ef\ (which is a fraction of the
\Bf, $E=\eta B$), one can find {\it the maximum energy of curvature emission
within the Crab pulsar \ms}:
$
\epsilon_{br} 
%(3 \pi)^{7/4}  { \hbar \over ( c e) ^{3/4} } \eta^{3/4} \sqrt{\xi} \,  {  B_{NS}^{3/4} R_{NS}^{9/4}\over P^{7/4}}
 =
(150\, {\rm GeV}) \, \eta^{3/4} \sqrt{\xi} .
% = 5\, {\rm GeV} \, \eta_{-2} ^{3/4} \sqrt{\xi} 
% \label{1}
$ \citep{2012ApJ...754...33L}.
The  possibility  that the emission above the break energy is produced  as a tail to
the curvature emission  is excluded by the fact that 
the spectrum of the Crab pulsar reported by  MAGIC and VERITAS {\it does not show the
exponential cut-off} indicative of radiation reaction-limited curvature emission. Phase-dependent variations of the break energy  at few GeVs should not affect the shape of the spectrum at
sufficiently high energies, \eg above 10\,GeV.

Importantly, it is the high-energy tails that carry information about the emission mechanism and thus deserve
special consideration. The two competing models of pulsar $\gamma$-ray emission predict
qualitatively different spectra well above the break energy (curvature emission predicts
an exponential cut-off, while inverse-Compton scattering predicts a power-law).

One of the problems of the broadband double power-law fits,  commonly used for a typical {\it Fermi}-LAT
spectrum of a bright $\gamma$-ray pulsar, is that  most of the errors accumulate due to the  {\it
arbitrary} parametrization of the spectral near the peak, where the two power laws connect.  A different parametrization of this spectral part
 can produce substantially different fits~\citep{2012ApJ...757...88L}. 

\cite{2013MNRAS.431.2580L} showed 
that  the broadband  spectrum of the Crab pulsar from UV to very high-energy
$\gamma$-rays---nearly ten decades in energy---can be reproduced  within the framework of 
the
cyclotron self-Compton model \citep[see also][]{2015ApJ...811...63H}. Emission is produced  by two counter-streaming beams
within
the outer gaps, at distances above $\sim$20 NS radii. The outward moving beam  produces
UV--X-ray photons via Doppler-boosted cyclotron emission, and GeV photons by
inverse-Compton
scattering the cyclotron photons produced by the inward going beam. The scattering occurs
deep in the Klein-Nishina regime, whereby the  IC component provides a direct
measurement
of the particle distribution within the magnetosphere.   The required plasma multiplicity
is
high, on the order of $10^6$ to $10^7$, but is consistent with the average particle flux injected into
the
pulsar wind nebula. Klein-Nishina reduction in the scattering cross-section (and the
corresponding  reduction
of the electron energy loss rate)  allows the primary  leptons to be accelerated to very
high energies with hard spectra.  The secondary plasma is less energetic but more dense
and has approximately the same energy content as the primary beam, producing
cyclotron self-Compton radiation. The  cyclotron component has a broad peak in the
UV--X-ray range, while the IC component extends to hundreds of GeV.  Below $\sim$1\,GeV
the  curvature emission from the primary beam can contribute a substantial flux fraction.
The IC emission from the primary beam extends well into the TeV regime but has proven
difficult to detect by Cherenkov telescopes. 
% careful about self-plagiarism here
% cyclotron vs. synchrotron? 
% ``Doppler-booster''?

How special is the Crab in terms of the high-energy power-law tail?
\cite{2012ApJ...757...88L} analyzed  the high energy emission from Geminga (using
post-processed data)  and concluded that the high energy part of the spectrum is better
fit with a power-law shape than with an exponential cut-off.  The non-detection by
VERITAS~\citep{2015ApJ...800...61A} is consistent with the fairly steep power-law slope
inferred
by \cite{2012ApJ...757...88L}.  \cite{2014ApJ...797L..13L} have reported the
detection of pulsed emission from Vela $> 50$\,GeV, noting that the observed
spectrum above 10\,GeV of the
Vela pulsar is harder than a simple exponential function predicted in the exponential
cut-off spectral shape of the tail.
\cite{2015ApJ...804...86M} has performed a statistical analysis of high
energy tails using a stacked analysis of 115 {\it Fermi}-LAT-detected
pulsars, finding no significant stacked excess at energies above 50\,GeV.

In this paper we present the results of a complementary analysis of high-energy pulsar
tails, defined as emission above 10\,GeV.  We chose  four bright $\gamma$-ray pulsars: PSR
J0007+7303, PSR J0633+1746 (Geminga), PSR J0835--4510 (Vela) and  PSR J2021+3651,
and we perform detailed spectral fits for each pulsar.

\section{Fermi-LAT Data Analysis \& Results}

The {\it Fermi}-LAT is an electron-positron pair-conversion telescope sensitive to gamma-ray photons with energies 
between 20\,MeV and 300\,GeV.  It has a FoV of $\sim2.5$\,sr and attains full-sky coverage approximately every 
three hours.  For a complete description of the instrument, see~\cite{2009ApJ...697.1071A,2012ApJS..203....4A}.

To obtain the spectral parameters for each of the four pulsars, the {\it Fermi}-LAT \texttt{Science Tools 
v10r0p5} with Pass 8 reprocessed instrument response functions and the standard quality cuts described 
in~\cite{2012ApJS..199...31N} are used. Roughly 7\,yr of ``source''-class events with energies between 100\,MeV 
and 100\,GeV collected between 4 August 2008 and 12 August 2015 within a $20^\circ$ region-of-interest (ROI) of 
the 2PC location of each pulsar are processed with the maximum likelihood fitting routine.  

The spectral analysis for each pulsar except Vela
presented here follows the same steps outlined in the second {\it Fermi}-LAT pulsar
catalog~\citep[2PC][]{2013ApJS..208...17A}. 
To generate spectral energy distributions of the {\it Fermi}-LAT data for each pulsar, a
binned maximum likelihood 
analysis is done in logarithmically spaced energy bins spanning 
the range 100\,MeV to 100\,GeV. Spectral models 
for all sources in the 3FGL catalog~\citep{2015ApJS..218...23A} in the ROI in addition to the galactic and 
isotropic diffuse backgrounds (gll\_iem\_v06.fits, iso\_P8R2\_SOURCE\_V6\_v06.txt) are included in the likelihood 
fitting.  Sources within a circle $4^{\circ}$ in radius are modeled as power laws of the form 
\begin{equation}
 {dN \over dE} = N_0 {\left(E \over E_0\right)}^{\gamma},
\end{equation} 
where $N_0$ is the flux normalization, $E_0$ is fixed to 300\,GeV, and $\gamma$ is the spectral index.
These sources have only their normalization parameter left free in the fitting 
routine.  All other parameters are fixed except for the normalization on the galactic and isotropic diffuse 
background models.  In each bin, the pulsar is modeled as a point source with a simple power-law spectral shape 
with a fixed spectral index of $-2$.  An iterative spectral binning is performed starting
at three bins per decade in energy and increasing to four bins per decade, five, and so
on. The spectrum containing the greatest number of bins above 10\,GeV is kept for
subsequent fitting, while the others are discarded.

The {\it Fermi}-LAT data analysis for the Vela pulsar is done differently than described above due to three nearby 
bright diffuse sources (Vela X, Vela Jr., and Puppis A) that must be taken into account in order to achieve the 
greatest sensitivity.  To extract spectral parameters for the diffuse sources, the data are first phase-folded 
with the the \texttt{Tempo2} pulsar timing package~\citep{2006MNRAS.369..655H} using a publicly available timing 
solution from a webpage maintained by M. 
Kerr\footnote{\texttt{\detokenize{www.slac.stanford.edu/~kerrm/fermi_pulsar_timing/}}}. 
Phase-folding the data allows 
selection of the off-pulse region to remove the contribution from the Vela pulsar, which allows more robust 
likelihood fitting for the nearby diffuse sources.   Since the timing solution is not valid for the full 7\,yr 
data set, only data recorded by the {\it Fermi}-LAT during the period of validity of the timing solution are used 
in the following analysis.  First, the off-pulse phase region between 0.8 -- 1.0 is selected to remove the Vela 
pulsar.  Subsequently, all normalization parameters in the model are scaled to 20\% of their nominal values, and 
the Vela pulsar model is removed.  All sources within $4^{\circ}$ of the Vela pulsar 2PC location are modeled as 
power laws with free normalization parameters, including the three diffuse sources and the galactic and isotropic 
diffuse components.  Likelihood analyses are done in logarithmically spaced energy bins to allow computation of 
the normalization parameters for each source in each bin.  The normalization parameters are then scaled by a 
factor of 5, and the spectral model for the Vela pulsar is added.  The normalization parameters for the three 
diffuse sources are then fixed to the previously computed values in each energy bin, and the Vela pulsar SED is 
generated in the same way as described in the previous paragraph.       

\begin{figure*}[h]
\center
\epsscale{1.0}
\includegraphics[width=6.6in]{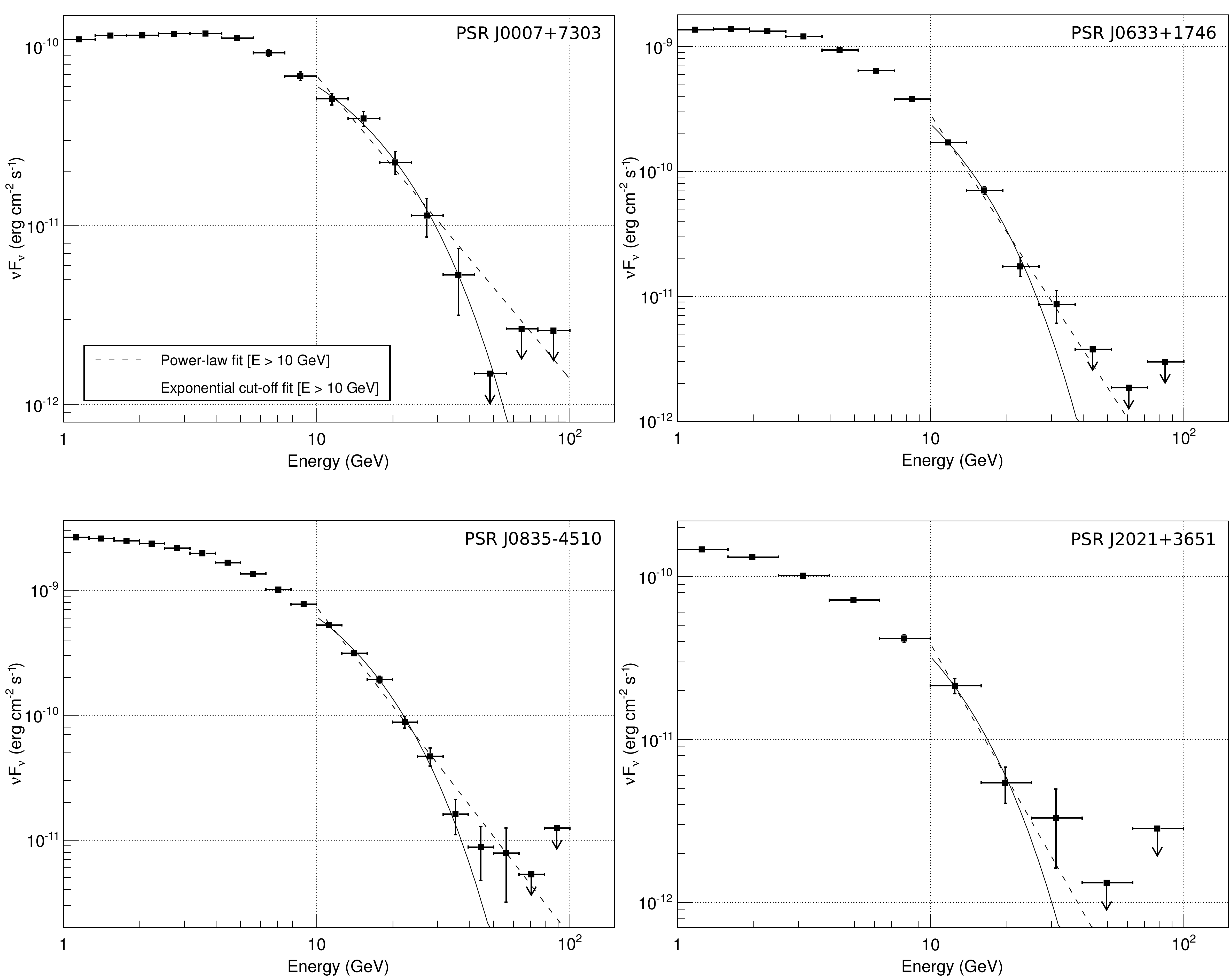}
\caption{ Spectral energy distributions derived from the {\it Fermi}-LAT data for the four
pulsars: PSR J0007+7303, PSR J0633+1746 (Geminga), PSR J0835--4510 (Vela), and PSR
J2021+3651. In each panel, the solid line shows the power-law fit, while the dashed line
shows the exponential cut-off fit. Each fit is applied above 10\,GeV. \label{seds}}
\end{figure*}

Phase-averaged spectral energy distributions for each pulsar are shown in
Figure~\ref{seds}.  Each SED is fitted 
with both a power-law and an exponential cut-off above 10\,GeV to test the preferred
spectral shape.  The choice of 10\,GeV is an arbitrary starting point for probing
 the highest energies of the spectra.  However, 10\,GeV is above a few times the
break energy for each pulsar reported in the 2PC, which helps to reduce contamination by
the harder portion of the spectra at lower energies.  Spectral 
indices for 
the power-law fits and fit probabilities are summarized in Table~\ref{tab:fit_summary}.   
\newline
\newcolumntype{C}{>{\centering\arraybackslash}X}
\begin{table*}[h]
\centering
\begin{tabularx}{1.0\textwidth}{CCCCC}
  \hline
  Pulsar & Power-law spectral index above 10\,GeV & Power-law fit chi-square / ndf (probability) & Exponential 
cut-off fit chi-square / ndf (probability) & Notes
\\ \hline
  PSR J0007+7303 & $-3.69 \pm 0.016$ & 5.26 / 3 (0.15) & 0.95 / 3 (0.81) & \\
  PSR J0633+1746 & $-5.12 \pm 0.16$ & 5.54 / 2 (0.06) & 4.87 / 2 (0.09) & Geminga \\ 
  PSR J0835--4510 & $-4.54 \pm 0.08$ & 14.1 / 5 (0.02) & 8.06 / 5 (0.15) & Vela \\
  PSR J2021+3651 & $-4.73 \pm 0.54$ & 1.10 / 1 (0.29) & 2.40 / 1 (0.12) & \\
  \hline
\end{tabularx}
\caption{ Spectral indices for the power-law fits, reduced chi-squares, and fit probabilities for the SED of each pulsar. }
\label{tab:fit_summary}
\end{table*}

The fit probabilities for either the power-law fit or the fit with an exponential cut-off are comparable. Therefore, one has to conclude that both models describe the data equally well and,
therefore, both models are equally preferred. The probability of a fit to the Vela data with a power law is the worst of all four power law fits. In the case of Vela, the data points at
the lowest energies included in the fit contribute the most to the $\chi^2$. In that energy range the spectrum shows clear curvature, highlighting the problem of arbitrarily
choosing 10\,GeV as the energy threshold for the fit. It also emphasizes that the sensitivity of the method strongly depends on how far away from the break in the spectrum the fit
begins.

In order to alleviate the subjective choice of where to start the fit, the spectra are also fitted with a power law plus sub-exponential cutoff (PLSEC) of the form
\begin{equation}
 {dN \over dE} = N_0 {\left(E \over E_0\right)}^{\gamma_1} \textrm{exp}\left(-{\left(E \over E_c\right)}^{\gamma_2}\right),
\end{equation}
where $E_0$ is fixed to 300\,GeV, and $N_0$, $E_c$, $\gamma_1 $ and $\gamma_2$ are free parameters, and a smooth, broken power
law (SBPL) of the form
\begin{equation}
 {dN \over dE} = N_0 {\left(E \over E_0\right)}^{\gamma_1}\left( 1 + {\left(E \over E_b	\right)}^{\gamma_1 - \gamma_2 \over \beta} \right)^{-\beta},
\end{equation}
where $E_0$ is again fixed to 300\,GeV, and $N_0$, $\gamma_1$, $\gamma_2$, $E_b$, and $\beta$ are all left free in the fit. 
%Two strategies are attempted. In the first, 
The parameters of the fit functions are constrained by fitting them to
data points below 10\,GeV. For both the PLSEC and the SBPL, including the minimum number of data points below 10\,GeV to constrain the parameters of the fit yields acceptable fit
probabilities. However, including more data points at lower energies in the fits, the fits fail with probabilities quickly sinking below $10^{-5}$. 
%which cannot be considered acceptable anymore. 
For all energy ranges for the fits, the PLSEC clearly under-predicts the flux above 10\,GeV, and the SBPL law clearly over-predicts the
data---neither of the fitted functions match the data above 10\,GeV. 

%Our second strategy was to fit the SBPL and the PLSEC to the data points over an energy range around where the spectra start to turnover,
%\emph{i.e.}\, between 1 and 10\,GeV. In these cases the fits yield acceptable fit probabilities comparable with our results where we only fit data points above 10\,GeV.

We note that the upper limit at 48.4\,GeV for PSR J0007+7303 is a factor of $\sim$3.5 below the power-law fit, but it still lies within 1.2\,$\sigma$ of the fit taking the error on the fit
into account. 
%of the bow-tie of the power-law fit.
Therefore, the upper limit at 48.4\,GeV is not inconsistent with the power-law scenario.

\section{Discussion}

We have performed an analysis of the high energy tails, defined as emission above 10\,GeV,
of four of the brightest {\it Fermi}-LAT-detected $\gamma$-ray pulsars. Inn our
approach, the binning was adjusted iteratively starting at three bins per decade
and increasing, keeping the spectrum with the most bins above 10\,GeV for subsequent
fitting. In the second {\it Fermi}-LAT pulsar
catalog~\citep[2PC][]{2013ApJS..208...17A} the binning was arbitrary, mostly tailored to
weaker sources. As a result, for the bright sources considered in this work, the binning
was not optimal.

The fit results for all four of the pulsars analyzed in this study do not allow any
firm claim  to be made about the shape of the spectra above 10\,GeV.  In no case is the
power-law fit or exponential cut-off fit significantly preferred over the other, though
the exponential cut-off is marginally preferred in the case of PSR J0007+7303 (which is
the dimmest pulsar of all considered in this work). 
\iffalse
Previously, in the case of Geminga, 
\cite{2012ApJ...757...88L} used post-processed data, with the standard binning, while in
the present work binning was adjusted to produce maximum likelihood of a fit. 
\fi
% I'm not sure this is needed here

One of the unquantified biases in our analysis is the choice of the lower energy cut-off,
taken here to be 10\,GeV. This choice  is motivated by our attempt to avoid fitting the
roll-off region near the break energies. 
% Again I think ``roll-off'' is too colloquial.

\iffalse
Alternatively we chose the lower cut at three
times the peak energy. In the sample, PSR J0007+7303 has the highest peak energy of  $4.7$
GeV and is the weakest of all, so a choice of higher energy cut reduces the statistical
significance.   The break energies in Vela and PSR J2021+3651  are  both  $3.0$ GeV.  The
only exception is Geminga, with the break energy of $2.2$ GeV. 
\fi
% I think this part just adds confusion...  We opted not to fit @ 3x Eb since 3x Eb is
% very close to (or > than) 10 GeV anyway for all 4 except J0007. 

Our work leaves open the question of the shape of the high-energy gamma-ray spectral
tails of these four pulsars. 

\acknowledgments

\bibliographystyle{apj}  
%\bibliography{./BibTex}
 \bibliography{/Users/maxim/Home/Research/BibTex} 

\begin{thebibliography}{25}
\expandafter\ifx\csname natexlab\endcsname\relax\def\natexlab#1{#1}\fi

\bibitem[{{Abdo} {et~al.}(2013){Abdo}, {Ajello}, {Allafort}, {Baldini},
  {Ballet}, {Barbiellini}, {Baring}, {Bastieri}, {Belfiore}, {Bellazzini}, \&
  et~al.}]{2013ApJS..208...17A}
{Abdo}, A.~A., {Ajello}, M., {Allafort}, A., {et~al.} 2013, \apjs, 208, 17

\bibitem[{{Acero} {et~al.}(2015){Acero}, {Ackermann}, {Ajello}, {Albert},
  {Atwood}, {Axelsson}, {Baldini}, {Ballet}, {Barbiellini}, {Bastieri},
  {Belfiore}, {Bellazzini}, {Bissaldi}, {Blandford}, {Bloom}, {Bogart},
  {Bonino}, {Bottacini}, {Bregeon}, {Britto}, {Bruel}, {Buehler}, {Burnett},
  {Buson}, {Caliandro}, {Cameron}, {Caputo}, {Caragiulo}, {Caraveo},
  {Casandjian}, {Cavazzuti}, {Charles}, {Chaves}, {Chekhtman}, {Cheung},
  {Chiang}, {Chiaro}, {Ciprini}, {Claus}, {Cohen-Tanugi}, {Cominsky}, {Conrad},
  {Cutini}, {D'Ammando}, {de Angelis}, {DeKlotz}, {de Palma}, {Desiante},
  {Digel}, {Di Venere}, {Drell}, {Dubois}, {Dumora}, {Favuzzi}, {Fegan},
  {Ferrara}, {Finke}, {Franckowiak}, {Fukazawa}, {Funk}, {Fusco}, {Gargano},
  {Gasparrini}, {Giebels}, {Giglietto}, {Giommi}, {Giordano}, {Giroletti},
  {Glanzman}, {Godfrey}, {Grenier}, {Grondin}, {Grove}, {Guillemot}, {Guiriec},
  {Hadasch}, {Harding}, {Hays}, {Hewitt}, {Hill}, {Horan}, {Iafrate}, {Jogler},
  {J{\'o}hannesson}, {Johnson}, {Johnson}, {Johnson}, {Johnson}, {Kamae},
  {Kataoka}, {Katsuta}, {Kuss}, {La Mura}, {Landriu}, {Larsson}, {Latronico},
  {Lemoine-Goumard}, {Li}, {Li}, {Longo}, {Loparco}, {Lott}, {Lovellette},
  {Lubrano}, {Madejski}, {Massaro}, {Mayer}, {Mazziotta}, {McEnery},
  {Michelson}, {Mirabal}, {Mizuno}, {Moiseev}, {Mongelli}, {Monzani},
  {Morselli}, {Moskalenko}, {Murgia}, {Nuss}, {Ohno}, {Ohsugi}, {Omodei},
  {Orienti}, {Orlando}, {Ormes}, {Paneque}, {Panetta}, {Perkins},
  {Pesce-Rollins}, {Piron}, {Pivato}, {Porter}, {Racusin}, {Rando}, {Razzano},
  {Razzaque}, {Reimer}, {Reimer}, {Reposeur}, {Rochester}, {Romani},
  {Salvetti}, {S{\'a}nchez-Conde}, {Saz Parkinson}, {Schulz}, {Siskind},
  {Smith}, {Spada}, {Spandre}, {Spinelli}, {Stephens}, {Strong}, {Suson},
  {Takahashi}, {Takahashi}, {Tanaka}, {Thayer}, {Thayer}, {Thompson},
  {Tibaldo}, {Tibolla}, {Torres}, {Torresi}, {Tosti}, {Troja}, {Van Klaveren},
  {Vianello}, {Winer}, {Wood}, {Wood}, {Zimmer}, \& {Fermi-LAT
  Collaboration}}]{2015ApJS..218...23A}
{Acero}, F., {Ackermann}, M., {Ajello}, M., {et~al.} 2015, \apjs, 218, 23

\bibitem[{{Ackermann} {et~al.}(2012){Ackermann}, {Ajello}, {Albert},
  {Allafort}, {Atwood}, {Axelsson}, {Baldini}, {Ballet}, {Barbiellini},
  {Bastieri}, {Bechtol}, {Bellazzini}, {Bissaldi}, {Blandford}, {Bloom},
  {Bogart}, {Bonamente}, {Borgland}, {Bottacini}, {Bouvier}, {Brandt},
  {Bregeon}, {Brigida}, {Bruel}, {Buehler}, {Burnett}, {Buson}, {Caliandro},
  {Cameron}, {Caraveo}, {Casandjian}, {Cavazzuti}, {Cecchi}, {{\c C}elik},
  {Charles}, {Chaves}, {Chekhtman}, {Cheung}, {Chiang}, {Ciprini}, {Claus},
  {Cohen-Tanugi}, {Conrad}, {Corbet}, {Cutini}, {D'Ammando}, {Davis}, {de
  Angelis}, {DeKlotz}, {de Palma}, {Dermer}, {Digel}, {Silva}, {Drell},
  {Drlica-Wagner}, {Dubois}, {Favuzzi}, {Fegan}, {Ferrara}, {Focke}, {Fortin},
  {Fukazawa}, {Funk}, {Fusco}, {Gargano}, {Gasparrini}, {Gehrels}, {Giebels},
  {Giglietto}, {Giordano}, {Giroletti}, {Glanzman}, {Godfrey}, {Grenier},
  {Grove}, {Guiriec}, {Hadasch}, {Hayashida}, {Hays}, {Horan}, {Hou}, {Hughes},
  {Jackson}, {Jogler}, {J{\'o}hannesson}, {Johnson}, {Johnson}, {Johnson},
  {Kamae}, {Katagiri}, {Kataoka}, {Kerr}, {Kn{\"o}dlseder}, {Kuss}, {Lande},
  {Larsson}, {Latronico}, {Lavalley}, {Lemoine-Goumard}, {Longo}, {Loparco},
  {Lott}, {Lovellette}, {Lubrano}, {Mazziotta}, {McConville}, {McEnery},
  {Mehault}, {Michelson}, {Mitthumsiri}, {Mizuno}, {Moiseev}, {Monte},
  {Monzani}, {Morselli}, {Moskalenko}, {Murgia}, {Naumann-Godo}, {Nemmen},
  {Nishino}, {Norris}, {Nuss}, {Ohno}, {Ohsugi}, {Okumura}, {Omodei},
  {Orienti}, {Orlando}, {Ormes}, {Paneque}, {Panetta}, {Perkins},
  {Pesce-Rollins}, {Pierbattista}, {Piron}, {Pivato}, {Porter}, {Racusin},
  {Rain{\`o}}, {Rando}, {Razzano}, {Razzaque}, {Reimer}, {Reimer}, {Reposeur},
  {Reyes}, {Ritz}, {Rochester}, {Romoli}, {Roth}, {Sadrozinski}, {Sanchez},
  {Saz Parkinson}, {Sbarra}, {Scargle}, {Sgr{\`o}}, {Siegal-Gaskins},
  {Siskind}, {Spandre}, {Spinelli}, {Stephens}, {Suson}, {Tajima}, {Takahashi},
  {Tanaka}, {Thayer}, {Thayer}, {Thompson}, {Tibaldo}, {Tinivella}, {Tosti},
  {Troja}, {Usher}, {Vandenbroucke}, {Van Klaveren}, {Vasileiou}, {Vianello},
  {Vitale}, {Waite}, {Wallace}, {Winer}, {Wood}, {Wood}, {Wood}, {Yang}, \&
  {Zimmer}}]{2012ApJS..203....4A}
{Ackermann}, M., {Ajello}, M., {Albert}, A., {et~al.} 2012, \apjs, 203, 4

\bibitem[{{Aliu} {et~al.}(2008){Aliu}, {Anderhub}, {Antonelli}, {Antoranz},
  {Backes}, {Baixeras}, {Barrio}, {Bartko}, {Bastieri}, {Becker}, {Bednarek},
  {Berger}, {Bernardini}, {Bigongiari}, {Biland}, {Bock}, {Bonnoli}, {Bordas},
  {Bosch-Ramon}, {Bretz}, {Britvitch}, {Camara}, {Carmona}, {Chilingarian},
  {Commichau}, {Contreras}, {Cortina}, {Costado}, {Covino}, {Curtef}, {Dazzi},
  {De Angelis}, {De Cea del Pozo}, {de los Reyes}, {De Lotto}, {De Maria}, {De
  Sabata}, {Delgado Mendez}, {Dominguez}, {Dorner}, {Doro}, {Els{\"a}sser},
  {Errando}, {Fagiolini}, {Ferenc}, {Fernandez}, {Firpo}, {Fonseca}, {Font},
  {Galante}, {Garcia Lopez}, {Garczarczyk}, {Gaug}, {Goebel}, {Hadasch},
  {Hayashida}, {Herrero}, {H{\"o}hne}, {Hose}, {Hsu}, {Huber}, {Jogler},
  {Kranich}, {La Barbera}, {Laille}, {Leonardo}, {Lindfors}, {Lombardi},
  {Longo}, {Lopez}, {Lorenz}, {Majumdar}, {Maneva}, {Mankuzhiyil}, {Mannheim},
  {Maraschi}, {Mariotti}, {Martinez}, {Mazin}, {Meucci}, {Meyer}, {Miranda},
  {Mirzoyan}, {Moles}, {Moralejo}, {Nieto}, {Nilsson}, {Ninkovic}, {Otte},
  {Oya}, {Paoletti}, {Paredes}, {Pasanen}, {Pascoli}, {Pauss}, {Pegna},
  {Perez-Torres}, {Persic}, {Peruzzo}, {Piccioli}, {Prada}, {Prandini},
  {Puchades}, {Raymers}, {Rhode}, {Rib{\'o}}, {Rico}, {Rissi}, {Robert},
  {R{\"u}gamer}, {Saggion}, {Saito}, {Salvati}, {Sanchez-Conde}, {Sartori},
  {Satalecka}, {Scalzotto}, {Scapin}, {Schweizer}, {Shayduk}, {Shinozaki},
  {Shore}, {Sidro}, {Sierpowska-Bartosik}, {Sillanp{\"a}{\"a}}, {Sobczynska},
  {Spanier}, {Stamerra}, {Stark}, {Takalo}, {Tavecchio}, {Temnikov}, {Tescaro},
  {Teshima}, {Tluczykont}, {Torres}, {Turini}, {Vankov}, {Venturini}, {Vitale},
  {Wagner}, {Wittek}, {Zabalza}, {Zandanel}, {Zanin}, {Zapatero}, {de Jager},
  {de Ona Wilhelmi}, \& {MAGIC Collaboration}}]{2008Sci...322.1221A}
{Aliu}, E., {Anderhub}, H., {Antonelli}, L.~A., {et~al.} 2008, Science, 322,
  1221

\bibitem[{{Aliu} {et~al.}(2015){Aliu}, {Archambault}, {Archer}, {Aune},
  {Barnacka}, {Beilicke}, {Benbow}, {Bird}, {Buckley}, {Bugaev}, {Byrum},
  {Cardenzana}, {Cerruti}, {Chen}, {Ciupik}, {Connolly}, {Cui}, {Dickinson},
  {Dumm}, {Eisch}, {Errando}, {Falcone}, {Feng}, {Finley}, {Fleischhack},
  {Fortin}, {Fortson}, {Furniss}, {Gillanders}, {Griffin}, {Griffiths},
  {Grube}, {Gyuk}, {H{\aa}kansson}, {Hanna}, {Holder}, {Humensky}, {Johnson},
  {Kaaret}, {Kar}, {Kertzman}, {Kieda}, {Krennrich}, {Kumar}, {Lang},
  {Lyutikov}, {Madhavan}, {Maier}, {McArthur}, {McCann}, {Meagher}, {Millis},
  {Moriarty}, {Mukherjee}, {Nieto}, {O'Faol{\'a}in de Bhr{\'o}ithe}, {Ong},
  {Otte}, {Park}, {Pohl}, {Popkow}, {Prokoph}, {Pueschel}, {Quinn}, {Ragan},
  {Reyes}, {Reynolds}, {Richards}, {Roache}, {Santander}, {Sembroski},
  {Shahinyan}, {Smith}, {Staszak}, {Telezhinsky}, {Tucci}, {Tyler}, {Varlotta},
  {Vincent}, {Wakely}, {Weinstein}, {Williams}, {Zajczyk}, \&
  {Zitzer}}]{2015ApJ...800...61A}
{Aliu}, E., {Archambault}, S., {Archer}, A., {et~al.} 2015, \apj, 800, 61

\bibitem[{{Ansoldi} {et~al.}(2016){Ansoldi}, {Antonelli}, {Antoranz}, {Babic},
  {Bangale}, {Barres de Almeida}, {Barrio}, {Becerra Gonz{\'a}lez}, {Bednarek},
  {Bernardini}, {Biasuzzi}, {Biland}, {Blanch}, {Bonnefoy}, {Bonnoli},
  {Borracci}, {Bretz}, {Carmona}, {Carosi}, {Colin}, {Colombo}, {Contreras},
  {Cortina}, {Covino}, {Da Vela}, {Dazzi}, {De Angelis}, {De Caneva}, {De
  Lotto}, {de O{\~n}a Wilhelmi}, {Delgado Mendez}, {Di Pierro}, {Dominis
  Prester}, {Dorner}, {Doro}, {Einecke}, {Eisenacher Glawion}, {Elsaesser},
  {Fern{\'a}ndez-Barral}, {Fidalgo}, {Fonseca}, {Font}, {Frantzen}, {Fruck},
  {Galindo}, {Garc{\'{\i}}a L{\'o}pez}, {Garczarczyk}, {Garrido Terrats},
  {Gaug}, {Godinovi{\'c}}, {Gonz{\'a}lez Mu{\~n}oz}, {Gozzini}, {Hanabata},
  {Hayashida}, {Herrera}, {Hirotani}, {Hose}, {Hrupec}, {Hughes}, {Idec},
  {Kellermann}, {Knoetig}, {Kodani}, {Konno}, {Krause}, {Kubo}, {Kushida}, {La
  Barbera}, {Lelas}, {Lewandowska}, {Lindfors}, {Lombardi}, {Longo},
  {L{\'o}pez}, {L{\'o}pez-Coto}, {L{\'o}pez-Oramas}, {Lorenz}, {Makariev},
  {Mallot}, {Maneva}, {Mannheim}, {Maraschi}, {Marcote}, {Mariotti},
  {Mart{\'{\i}}nez}, {Mazin}, {Menzel}, {Miranda}, {Mirzoyan}, {Moralejo},
  {Munar-Adrover}, {Nakajima}, {Neustroev}, {Niedzwiecki}, {Nevas Rosillo},
  {Nilsson}, {Nishijima}, {Noda}, {Orito}, {Overkemping}, {Paiano},
  {Palatiello}, {Paneque}, {Paoletti}, {Paredes}, {Paredes-Fortuny}, {Persic},
  {Poutanen}, {Prada Moroni}, {Prandini}, {Puljak}, {Reinthal}, {Rhode},
  {Rib{\'o}}, {Rico}, {Rodriguez Garcia}, {Saito}, {Saito}, {Satalecka},
  {Scalzotto}, {Scapin}, {Schultz}, {Schweizer}, {Shore}, {Sillanp{\"a}{\"a}},
  {Sitarek}, {Snidaric}, {Sobczynska}, {Stamerra}, {Steinbring}, {Strzys},
  {Takalo}, {Takami}, {Tavecchio}, {Temnikov}, {Terzi{\'c}}, {Tescaro},
  {Teshima}, {Thaele}, {Torres}, {Toyama}, {Treves}, {Ward}, {Will}, \&
  {Zanin}}]{2016A&A...585A.133A}
{Ansoldi}, S., {Antonelli}, L.~A., {Antoranz}, P., {et~al.} 2016, \aap, 585,
  A133

\bibitem[{{Atwood} {et~al.}(2009){Atwood}, {Abdo}, {Ackermann}, {Althouse},
  {Anderson}, {Axelsson}, {Baldini}, {Ballet}, {Band}, {Barbiellini}, \&
  et~al.}]{2009ApJ...697.1071A}
{Atwood}, W.~B., {Abdo}, A.~A., {Ackermann}, M., {et~al.} 2009, \apj, 697, 1071

\bibitem[{{Bai} \& {Spitkovsky}(2010)}]{2010ApJ...715.1282B}
{Bai}, X.-N., \& {Spitkovsky}, A. 2010, \apj, 715, 1282

\bibitem[{{Baring} {et~al.}(1999){Baring}, {Ellison}, {Reynolds}, {Grenier}, \&
  {Goret}}]{Baring99}
{Baring}, M.~G., {Ellison}, D.~C., {Reynolds}, S.~P., {Grenier}, I.~A., \&
  {Goret}, P. 1999, \apj, 513, 311

\bibitem[{{Belyaev}(2015)}]{2015MNRAS.449.2759B}
{Belyaev}, M.~A. 2015, \mnras, 449, 2759

\bibitem[{{Cerutti} {et~al.}(2015){Cerutti}, {Philippov}, {Parfrey}, \&
  {Spitkovsky}}]{2015MNRAS.448..606C}
{Cerutti}, B., {Philippov}, A., {Parfrey}, K., \& {Spitkovsky}, A. 2015,
  \mnras, 448, 606

\bibitem[{{Chen} \& {Beloborodov}(2014)}]{2014ApJ...795L..22C}
{Chen}, A.~Y., \& {Beloborodov}, A.~M. 2014, \apjl, 795, L22

\bibitem[{{de Jager} {et~al.}(1996){de Jager}, {Harding}, {Michelson}, {Nel},
  {Nolan}, {Sreekumar}, \& {Thompson}}]{1996ApJ...457..253D}
{de Jager}, O.~C., {Harding}, A.~K., {Michelson}, P.~F., {et~al.} 1996, \apj,
  457, 253

\bibitem[{{Harding} \& {Kalapotharakos}(2015)}]{2015ApJ...811...63H}
{Harding}, A.~K., \& {Kalapotharakos}, C. 2015, \apj, 811, 63

\bibitem[{{Hobbs} {et~al.}(2006){Hobbs}, {Edwards}, \&
  {Manchester}}]{2006MNRAS.369..655H}
{Hobbs}, G.~B., {Edwards}, R.~T., \& {Manchester}, R.~N. 2006, MNRAS, 369, 655

\bibitem[{{Leung} {et~al.}(2014){Leung}, {Takata}, {Ng}, {Kong}, {Tam}, {Hui},
  \& {Cheng}}]{2014ApJ...797L..13L}
{Leung}, G.~C.~K., {Takata}, J., {Ng}, C.~W., {et~al.} 2014, \apjl, 797, L13

\bibitem[{{Lyutikov}(2010)}]{2010MNRAS.405.1809L}
{Lyutikov}, M. 2010, MNRAS, 405, 1809

\bibitem[{{Lyutikov}(2012)}]{2012ApJ...757...88L}
---. 2012, \apj, 757, 88

\bibitem[{{Lyutikov}(2013)}]{2013MNRAS.431.2580L}
---. 2013, \mnras, 431, 2580

\bibitem[{{Lyutikov} {et~al.}(2012){Lyutikov}, {Otte}, \&
  {McCann}}]{2012ApJ...754...33L}
{Lyutikov}, M., {Otte}, N., \& {McCann}, A. 2012, \apj, 754, 33

\bibitem[{{McCann}(2015)}]{2015ApJ...804...86M}
{McCann}, A. 2015, \apj, 804, 86

\bibitem[{{Nolan} {et~al.}(2012){Nolan}, {Abdo}, {Ackermann}, {Ajello},
  {Allafort}, {Antolini}, {Atwood}, {Axelsson}, {Baldini}, {Ballet}, \&
  et~al.}]{2012ApJS..199...31N}
{Nolan}, P.~L., {Abdo}, A.~A., {Ackermann}, M., {et~al.} 2012, \apjs, 199, 31

\bibitem[{{Romani}(1996)}]{1996ApJ...470..469R}
{Romani}, R.~W. 1996, \apj, 470, 469

\bibitem[{{Romani} \& {Yadigaroglu}(1995)}]{1995ApJ...438..314R}
{Romani}, R.~W., \& {Yadigaroglu}, I.-A. 1995, \apj, 438, 314

\bibitem[{{VERITAS Collaboration} {et~al.}(2011){VERITAS Collaboration},
  {Aliu}, {Arlen}, {Aune}, {Beilicke}, {Benbow}, {Bouvier}, {Bradbury},
  {Buckley}, {Bugaev}, {Byrum}, {Cannon}, {Cesarini}, {Christiansen}, {Ciupik},
  {Collins-Hughes}, {Connolly}, {Cui}, {Dickherber}, {Duke}, {Errando},
  {Falcone}, {Finley}, {Finnegan}, {Fortson}, {Furniss}, {Galante}, {Gall},
  {Gibbs}, {Gillanders}, {Godambe}, {Griffin}, {Grube}, {Guenette}, {Gyuk},
  {Hanna}, {Holder}, {Huan}, {Hughes}, {Hui}, {Humensky}, {Imran}, {Kaaret},
  {Karlsson}, {Kertzman}, {Kieda}, {Krawczynski}, {Krennrich}, {Lang},
  {Lyutikov}, {Madhavan}, {Maier}, {Majumdar}, {McArthur}, {McCann},
  {McCutcheon}, {Moriarty}, {Mukherjee}, {Nu{\~n}ez}, {Ong}, {Orr}, {Otte},
  {Park}, {Perkins}, {Pizlo}, {Pohl}, {Prokoph}, {Quinn}, {Ragan}, {Reyes},
  {Reynolds}, {Roache}, {Rose}, {Ruppel}, {Saxon}, {Schroedter}, {Sembroski},
  {{\c S}ent{\"u}rk}, {Smith}, {Staszak}, {Te{\v s}i{\'c}}, {Theiling},
  {Thibadeau}, {Tsurusaki}, {Tyler}, {Varlotta}, {Vassiliev}, {Vincent},
  {Vivier}, {Wakely}, {Ward}, {Weekes}, {Weinstein}, {Weisgarber}, {Williams},
  \& {Zitzer}}]{2011Sci...334...69V}
{VERITAS Collaboration}, {Aliu}, E., {Arlen}, T., {et~al.} 2011, Science, 334,
  69

\end{thebibliography}
\end{document}